\documentclass[10pt]{article}
\usepackage{latexsym}
\usepackage[dvips]{graphicx}
\usepackage{overcite}
\usepackage{amssymb}
\usepackage{amsbsy}
\usepackage{amsmath}
\usepackage{mathrsfs}
\usepackage{xr}
\usepackage{booktabs} 
\usepackage{multirow} 
\usepackage{xcolor}	  
\usepackage{mathtools}
\usepackage{graphicx}
\usepackage{tabularx}
\usepackage[labelformat=simple, font=small,labelfont=bf]{caption}
\usepackage{placeins}
\usepackage{authblk}
\graphicspath{ {./plot/} }

\DeclareRobustCommand*{\Figure}[3]{
   \begin{figure}[!htb]
   \begin{center}
   \noindent
   \includegraphics[width=#2]{#1}  
   \end{center}
   \caption{#3}
   \addtocontents{lof}{\vspace{\baselineskip}}
   \label{fig:#1}
   \end{figure}
}

\date{}
\newcommand{\be}{\begin{equation}}
\newcommand{\ee}{\end{equation}}

\begin{document}
\title{A Generic Explanation of the Mechanism of Co-solvency}

\author[1]{Xiangyu Zhang\footnote{E-mail: xzhan357@jh.edu}}
\author[2]{Dong Meng}
\affil[1]{Department of Chemical and Biomolecular Engineering, John Hopkins University, Baltimore, MD 21218, United States}
\affil[2]{Biomaterials Division, Department of Molecular Pathobiology, New York University, New York, NY 10010, United States}

\maketitle

\begin{abstract}

Polymer behavior in mixed solvents often exhibits intriguing phenomena, such as cosolvency, where a polymer that collapses in two individually poor solvents becomes soluble in their mixture. In this study, we employ a combination of theoretical modeling and computer simulations using a generic polymer solution model to uncover the underlying mechanisms driving this behavior. Moving beyond conventional explanations based on solvent–cosolvent immiscibility or chemistry-specific interactions, our findings highlight the critical role of mismatches in solvophobicity and solvation strength between the polymer and the two solvent components. We demonstrate that co-solvency arises from the interplay between van der Waals interactions and specific associations, such as hydrogen bonding. The bulk phase behavior of the solution is also examined, and the resulting phase diagram, calculated using Flory–Huggins theory, shows good agreement with experimental observations. This study offers a generalized  framework for understanding polymer cosolvency across diverse systems.

\end{abstract}

\section{Introduction}
The utilization of solvent mixture to induce polymer conformational transition is principle in functional polymeric materials processing \cite{bharadwaj2022cononsolvency,mukherji2020smart}. This covers a wide range of applications, reaching from catalysis \cite{kleinschmidt2020microgel}, hydrogel actuators \cite{wang2019multi} to self-assembled nano-structures \cite{kyriakos2014cononsolvency}. The mixture of two solvents cannot only turn the overall solvent quality worse, pertaining to cononsolvency phenomenon, but also can turn it better, corresponding to cosolvency phenomenon \cite{dudowicz2015communication}. It would be better to clarify some concepts about cosolvency phenomenon before moving further. Cosolvency means the improved solubility, but it has two different cases. The solubility of the solute increases monotonically or non-monotonically with the addition of the secondary solvent (cosolvent), corresponding to the maximum solubility at pure cosolvent or intermediate solvent composition, respectively. In areas like small molecules or lignin dissolution, monotonic change is commonly reported \cite{nayak2012solubility,soares2019hydrotropy}. In polymer system, cosolvency often refers to non-monotonic variation \cite{dudowicz2015communication}, and that is our focusing point in this study.  

\par
Cosolvency is a rare phenomenon and it has been less discussed compared with cononsolvency. Experimental results have shown that cosolvency is not only restricted to one specific type of polymer or one specific solvent/cosolvent pair. For example, water/alcohol (like methanol, propanol, t-butanol) \cite{cowie1987alcohol}, 1-chlorobutane/carbon tetrachloride \cite{katime1984polymer}, acetonitrile/1-chlorobutane \cite{horta1981preferential}, acetonitrile/1-butanol \cite{prolongo1984cosolvency} and alcohol/carbon tetrachloride \cite{katime1983polymer}, etc. \cite{nakata1988light, masegosa1984preferential,wolf1975measured,katime1985polymer+}, can all be cosolvency pairs for poly(methyl methacrylate) (PMMA). Except for PMMA, other polymer types, like polystyrene (PS), poly(vinyl-2-pyridine) (PV2P) and some others also exhibit cosolvency phenomenon in some solvent/cosolvent pairs \cite{cowie1983polymer,masegosa1987polystyrene,wolf1976pressure, palaiologou1988polymer,viras1988preferential}.  Besides, the enrichment of the cosolvent around the polymer coil or the preferential adsorption to the polymer is reported for some systems, such as methanol/carbon tetrachloride/PMMA, acetonitrile/1-butanol/PMMA, etc \cite{katime1984polymer,horta1981preferential,prolongo1984cosolvency, cowie1972polymer,viras1988preferential}. Generally speaking, the previous proposed mechanism can be categorized into two aspects.

\par
It has been proposed that cosolvency may arise from the solvent - cosolvent repulsive interactions in the framework of Flory-Huggins theory \cite{dudowicz2015communication}. The negative sign in front of $\chi_{SC}$ (solvent-cosolvent interaction strength) leads to opposite effect of the effective interaction  ($\chi_{eff}$) during varying $\chi_{SC}$ value. Thereby, the miscibility is promoted with the increase of the repulsive interaction between solvents and cosolvents \cite{dudowicz2015communication}. Similar argument has been made by Mukherji and his coworkers, though through a totally different path, stating that polymer swelling results from the depletion of density of solvents and cosolvents  \cite{mukherji2017depleted}. Their molecular dynamics (MD) simulation proves that the number density variation with the addition of the cosolvent results in the polymer swelling. In their study, the "dip" of the number density from the mean-field values arises from the incompatibility between solvents and cosolvents. In other words, the depletion of the number density of solvents and cosolvents actually is an effect of $\chi_{SC}$, of which the positive value is still the deciding factor in the mechanism.
\par
The above arguments basically ascribe the occurrence of cosolvency to the solvent/cosolvent repulsive interaction effect. But the picture becomes more intriguing with the introduction of association to the model. The association of solvents and cosolvents to the polymer is introduced to investigate the miscibility diagram in the framework of Flory-Huggins theory \cite{dudowicz2015phase}. In their conclusion, the $\chi_{SC}$ is still the deciding factor for the occurrence of cosolvency phenomenon when the association process of P-S and P-C have the same equilibrium constant, or in other words, same associating probability. The surprising point is that the $\chi_{SC}$ effect becomes less important when the equilibrium constant of P-S and P-C is distinct. And obviously, the phase diagram is more sensitive to the addition of the secondary solvents to the P/S binary mixture with P-S and P-C equilibrium constant being different. This may indicate that the picture is incomplete with ascribing cosolvency merely to $\chi_{SC}$ effect. Additionally, Among all types of cosolvency solvent/cosolvent pairs, some of their binary mixtures show the negative excess volume change, like alcohol/water \cite{shalmashi2014densities}, and some of them have positive excess volume change, like acetonitrile/1-butanl \cite{tahery2006density}. So, the number density change may not be the key point. Although most of solvent/cosolvent pairs of cosolvency show positive excess Gibbs free energy of mixing ($G_{E}$) \cite{villamanan1985excess,lama1965excess}, indicating a certain incompatibility, however, not all binary solvent mixture with positive $G_{E}$ can act as cosolvent pairs \cite{cowie1972polymer}.
\par
The other type of explanation proposes that cosolvent act as compatibilizer to dissolve the polymer in the solvent \cite{hoogenboom2008tuning, lambermont2010temperature, zhang2015polymers}.  PMMA cannot be well dissolved in alcohol, but it can adsorb water due to its strong hydrogen bonding accepting moieties, which can help polymer compatibilize with surrounding solvents and cosolvents \cite{zhang2015polymers}. Under this understanding, polymer-cosolvent contact is favored due to strong polymer-cosolvent interactions. Thereby, it can explain the shift of best miscibility ratio with the change of  solvent/cosolvent or polymer species, as the polarity is also changed. Further increase of cosolvent fraction will cause the formation of cosolvent clusters, and decreases the polymer solubility \cite{hoogenboom2008tuning}. But the question still lies here, the above explanation is a blurry statement. How to understand "compatibilize" effect? How many types of interactions are involved and how do they interact with each other? Obviously, we need a detailed explanation for it. To answer this question, an analytic theory and coarse-grained level simulation study are needed. In our study, Van Der Waals (VDW) type interaction caused by species polarity can be modeled by $\chi$ parameter, and the hydrogen bond is taken into account as reactive association. So, a simplified picture is built up based on the physical force. \par
In this manuscript, we first examine the impact of incorporating association into Flory–Huggins theory and its influence on the second-order virial coefficient under various parameter conditions. Next, we design three systems with distinct characteristics and perform Monte Carlo simulations to investigate the driving forces behind cosolvency. Finally, we compare the single-chain simulation results and bulk-phase theoretical predictions with experimental data, finding good agreement. This consistency indicates that the proposed mechanism can describe cosolvency phenomena at both the chain conformation and macroscopic phase levels.
\section{Method}
The system we study is homopolymer (P) with chain length being $N_{P}$ immersed in solvents (S) and cosolvents (C) mixture with both unit length.
\subsection{Associative Flory-Huggins Theory}
The way to deal with association is based on the method developed by A. Matsuyama and F. Tanaka \cite{matsuyama1990theory}. The "m-cluster" is formed by the association of one polymer chain with $m$ cosolvents. $n_{m+1}$ and $\phi_{m+1}$ represent the number and volume fraction of "m" cluster, respectively. $m$ being equal to zero indicates the pure polymer chain without any attached cosolvents. Solvents do not carry associating sites in the theory calculation. The Gibbs free energy of the system can be written as the sum of energy contributed by association, entropy and non-bonded interactions,
\begin{equation} \label{eq:7result_G_PSC}
\begin{aligned}
\frac{G}{k_{B}T}= & n_{0}\ln{\phi_{0}}+n_{S}\ln{\phi_{S}}+\sum_{m=0}^{f} n_{m+1}\ln{\phi_{m+1}} \\ 
& +\frac{z}{2}(\epsilon_{SS}n_{S}\phi_{S}+\epsilon_{CC}\phi_{C}(n_{0}+\sum_{m=0}^{f} n_{m+1}m)+\epsilon_{PP}\phi_{P}N_{P}\sum_{m=0}^{f}N_{m+1})\\ 
& +z(\epsilon_{PC}\phi_{C}N_{P}\sum_{m=0}^{f}n_{m+1}+\epsilon_{PS}\phi_{S}N_{P}\sum_{m=0}^{f}n_{m+1}+\epsilon_{SC}\phi_{S}(n_{0}+\sum_{m=0}^{f}n_{m+1}m)) \\ 
& +P(n_{0}+n_{S}+\sum_{m=0}^{f}n_{m+1}(m+n))+n_{0}\mu_{0}^{0}+\sum_{m=0}^{f}n_{m+1}\mu_{m+1}^{0} 
\end{aligned}
\end{equation} 
where $n_{0}$ is the number of unbounded cosolvents, $\phi_{0}$ is the volume fraction of unbounded cosolvents, $n_{C}$ is defined as the total number of cosolvents, $\phi_{C}$ is the overall volume fraction of cosolvents, $n_{S}$ is the number of solvents, $\phi_{S}$ is the volume fraction of solvents, $N_{P}$ is the polymer chain length, $n_{P}$ is total number of polymer chains, $\phi_{P}$ is the overall volume fraction of polymer segments, and therefore $\phi_{P}+\phi_{S}+\phi_{C}=1$, $f$ is the number of associative site each polymer chain, $\mu_{0}^{0}$ and $\mu_{m+1}^{0}$ are chemical potential of cosolvents in reference state and the isolated "m" cluster, $\epsilon_{XY}$ tells the interaction strength between "X" and "Y" component, $z$ is the coordination number and $P$ is the pressure of the system.\par
To find out osmotic pressure, the Gibbs free energy of solvents/cosolvents binary mixture (diffusible components) should be calculated,
\begin{equation} \label{eq:7result_G_SC}
\begin{aligned}
\frac{G^{0}}{k_{B}T}= & n_{C}^{0}\ln{\phi_{C}^{0}}+n_{S}^{0}\ln{\phi_{S}^{0}}+\frac{z}{2}\epsilon_{SS}n_{S}^{0}\phi_{S}^{0}+\frac{z}{2}\epsilon_{CC}n_{C}^{0}\phi_{C}^{0} \\ 
& +z\epsilon_{SC}n_{C}^{0}\phi_{C}^{0}+P^{0}(n_{S}^{0}+n_{C}^{0})+n_{C}^{0}\mu_{0}^{0}
\end{aligned}
\end{equation}
where $P^{0}$ denotes the pressure in the S/C mixture. The superscript $0$ indicates S/C binary mixture system. For example, $n_{S/C}^{0}$ means the number of cosolvents in S/C system. The reason why we need $\mu_{0}^{0}$ is that we treat cosolvents as reactive species requiring the use of it to describe the chemical potential of cosolvents in reference state, so it has the same physical meaning as $\mu_{0}^{0}$ in ternary system. \par
The chemical potential of $X$ species can be easily found by taking partial derivative of Gibbs free energy with respect to number of $X$ in the system. Hence, $\mu_{S}^{0}$, $\mu_{C}^{0}$ and $\mu_{S}$, $\mu_{C}$ can be obtained as following equations,
\be 
\frac{\mu_{S}^{0}}{k_{B}T}=(\frac{\partial G^{0}}{\partial n_{S}^{0}})_{n_{C}^{0},P^{0},T}=\ln{\phi_{S}^{0}}+\chi_{SC}{\phi_{C}^{0}}^{2}+P^{0}+\frac{z}{2}\epsilon_{SS}
\ee
\be 
\frac{\mu_{C}^{0}}{k_{B}T}=(\frac{\partial G^{0}}{\partial n_{C}^{0}})_{n_{S}^{0},P^{0},T}=\ln{\phi_{C}^{0}}+\chi_{SC}{\phi_{S}^{0}}^{2}+P^{0}+\frac{z}{2}\epsilon_{CC}+\mu_{0}^{0}
\ee
\begin{equation} 
\begin{aligned}
\frac{\mu_{S}}{k_{B}T} & = (\frac{\partial G}{\partial n_{S}})_{n_{C},n_{P},P,T} \\
& =-\phi_{0}+\ln{\phi_{S}}+1-\phi_{S}-\frac{\phi_{P}}{N_{P}}+\chi_{PS}\phi_{P}(1-\phi_{S})-\chi_{PC}\phi_{C}\phi_{P}+\chi_{SC}\phi_{C}(1-\phi_{S}) \\ 
& +\frac{z}{2}\epsilon_{SS}+P
\end{aligned}
\end{equation}
\begin{equation} 
\begin{aligned}
\frac{\mu_{C}}{k_{B}T} & = (\frac{\partial G}{\partial n_{0}})_{n_{S},n_{P},P,T} \\
& = \ln{\phi_{0}}+1-\phi_{0}-\phi_{S}-\frac{\phi_{P}}{N_{P}}+\chi_{PC}\phi_{P}(1-\phi_{C})-\chi_{PS}\phi_{S}\phi_{P}+\chi_{SC}\phi_{S}(1-\phi{C}) \\ 
& +\frac{z}{2}\epsilon_{CC}+P+\mu_{0}^{0}
\end{aligned}
\end{equation}
where $\chi_{XY} \equiv \frac{z}{2}(2\epsilon_{XY}-\epsilon_{XX}-\epsilon_{YY})$. \par
The osmotic pressure $\Pi$ is equal to $\Delta P \equiv P - P_{0}$, corresponding to the pressure difference in binary and ternary system, and chemical potential of diffusible component in each side should be equal. Hence, two $\Pi$ expressions can be obtained by equating the chemical potential of $S$ and $C$, named $\Pi(S)$ and $\Pi(C)$, respectively. 
\begin{equation}
\begin{aligned}
\Delta P = \Pi(S) & =  \ln{\phi_{S}^{0}}+\chi_{SC}{\phi_{C}^{0}}^{2}+\phi_{0}-\ln{\phi_{S}}-1+\phi_{S}+\frac{\phi_{P}}  
{N_{P}} \\ 
&-\chi_{PS}\phi_{P}(1-\phi_{S})+\chi_{PC}\phi_{P}\phi_{C}-\chi_{SC}\phi_{C}(1-\phi_{S})
\end{aligned}
\end{equation}
\begin{equation}
\begin{aligned}
\Delta P = \Pi(C) & =  \ln{\phi_{C}^{0}}+\chi_{SC}{\phi_{S}^{0}}^{2}+\phi_{0}-\ln{\phi_{0}}-1+\phi_{S}+\frac{\phi_{P}}{N_{P}} \\ 
& +\chi_{PS}\phi_{P}\phi_{S}-\chi_{PC}\phi_{P}(1-\phi_{C})-\chi_{SC}\phi_{S}(1-\phi_{C})
\end{aligned}
\end{equation}
Accordingly, two expressions are equal to each other, which means $\Pi(S)-\Pi(C)=0$.
\begin{equation} \label{eq:7result_PIS_PIC}
\ln{\frac{\phi_{S}^{0}}{1-\phi_{S}^{0}}}-2\chi_{SC}\phi_{S}^{0}+\ln{\frac{\phi_{0}}{\phi_{S}}}+(\chi_{PC}+\chi_{SC}-\chi_{PS})\phi_{P}+2\chi_{SC}\phi_{S}=0
\end{equation}
Because of the natural logarithm term in the equation, it cannot be solved directly. Hence, $\phi_{S}$ needs to be expanded in the term of $\phi_{P}$,
\begin{equation} \label{eq:7result_phiS}
\phi_{S}=\phi_{S}^{0}+A_{1}\phi_{P}+A_{2}\phi_{P}^{2}+... \ .
\end{equation}
Before moving further, we need to find an analytical expression for $\phi_{0}$. So, chemical reaction equilibrium is used, $\mu_{m+1}=\mu_{1}+m\mu_{C}$. 
\begin{equation}
\begin{aligned}
\frac{\mu_{m+1}}{k_{B}T} & = (\frac{\partial G}{\partial n_{m+1}})_{n_{S},n_{C},P,T} \\
& = -(m+n)\phi_{0}-(m+n)\phi_{S}+\ln{(\phi_{m+1})}+1-\frac{m+n}{N_{P}} \phi_{P}+\chi_{PS}\phi_{S}(N_{P}-(m+N_{P})\phi_{P}) \\
& +\chi_{PC}(m\phi_{P}+\phi_{C}(N_{P}-(m+N_{P})\phi_{P}))+\chi_{SC}\phi_{S}(m-(m+N_{P})\phi_{C}) \\
& + \mu_{m+1}^{0} + (m+N_{P})P + \frac{z}{2}(N_{P} \epsilon_{PP} + m\epsilon_{CC})
\end{aligned}
\end{equation}
\begin{equation}
\begin{aligned}
& \frac{\mu_{1}}{k_{B}T}=(\frac{\partial G}{\partial n_{1}})_{n_{S},n_{C},P,T} = 
 -N_{P}\phi_{0}-N_{P}\phi_{S}+\ln{\phi_1}+1-\phi_{P}+\chi_{PS}\phi_{S}(N_{P}-N_{P}\phi_{P})  \\
& +\chi_{PC}(\phi_{C}(N_{P}-N_{P}\phi_{P})) + \chi_{SC}(\phi_{S}(-N_{P}\phi_{C}))+\mu_{1}^{0}+N_{P}P+\frac{z}{2}N_{P}\epsilon_{PP}
\end{aligned}
\end{equation}
Substituting the above equations into the equilibrium condition, the following equation can be found, $\phi_{m+1}=K_{m}\phi_{1}\phi_{0}^{m}$, where $K_{m}=\exp{(m-E^{a})}$ and $E^{a}=-\mu_{1}^{0}-m\mu_{0}^{0}+\mu_{m+1}^{0}$. The way to calculate $E^{a}$ was reported by A. Matsuyama and F. Tanaka \cite{matsuyama1990theory}, so, derivation details will not be presented here. And it can be found that $K_{m}=\frac{m+n}{n}(\frac{f!}{m!(f+m)!})\lambda^{m}$, where $\lambda$ can be considered as association tendency, with $\lambda$ being larger indicating the increase of association rate. With knowing $K_m$ and $\phi = \sum \frac{N_{P}}{m+N_{P}} \phi_{m+1}$, it can be derived that,
\begin{equation}
\phi=\sum \frac{N_{P}}{m+N_{P}} \frac{m+n}{n}(\frac{f!}{m!(f+m)!})\lambda^{m}\phi_{1}\phi_{0}^{m}=\phi_{1}\sum (\frac{f!}{m!(f-m)!}) \lambda^{m}\phi_{0}^{m} .
\end{equation}
So, the following equation can be derived by using the relation $\sum_{k=0}^{n} \frac{a!}{b!(a-b)!} x^{b} = (1+x)^n$,  
\begin{equation}
\phi_{1} = \frac{\phi_{P}}{(1+\lambda \phi_{0})^{f}} .
\end{equation} 
$\phi_{1}$ can be substituted into mass conserved condition, $\phi_{0}+\phi_{S}+\sum\phi_{m+1}=1$, to solve the dependence of $\phi_{0}$ on $\phi_{P}$ and $\phi_{S}$,
\begin{equation} \label{eq:7result_phi0}
\begin{aligned}
& \Delta = (\lambda-1-(1+\frac{f}{n})\lambda\phi_{P}-\lambda\phi_{S})^{2}+4\lambda(1-\phi_{S}-\phi_{P}) \\
& \phi_{0}=\frac{1}{2\lambda}(\lambda-1-(1+\frac{f}{n})\lambda\phi_{P}-\lambda\phi_{S}+\sqrt{\Delta}) . 
\end{aligned}
\end{equation}
And, $\Delta$ is expanded about $\phi_{P}$ at $\phi_{P}\to 0$, 
\begin{equation}
\Delta\approx \frac{2\lambda^{3}\phi_{C}}{(\lambda\phi_{C}+1)^{3}}\phi_{P}^{2}+\frac{\lambda-\lambda^{2}\phi_{C}}{\lambda\phi_{C}+1}\phi_{P}+\lambda\phi_{C}+1 . 
\end{equation}
Next, let us move back to eq. \ref{eq:7result_PIS_PIC}, and try to solve $A_{1}$ and $A_{2}$ in eq. \ref{eq:7result_phiS}. By substituting eq. \ref{eq:7result_phi0} and eq. \ref{eq:7result_phiS} into eq. \ref{eq:7result_PIS_PIC}, and doing expansion of all $\ln$ term in the equation, it can be found that,
\begin{equation}
A_{1}=-\frac{\chi_{PC} + \chi_{SC} - \chi_{PS}- \frac{2\lambda \phi_{C}^{0}+1}{\phi_{C}^{0}(\lambda \phi_{C}^{0}+1)}}{2 \chi_{SC} + \frac{1}{\phi_{C}^{0}-1} - \frac{1}{\phi_{C}^{0}}}
\end{equation}
and $A_{2}$ will not be shown here, because of the lengthy expression. After obtaining $A_{1}$ and $A_{2}$, $\Pi\equiv \frac{1}{2}(\Pi(S)+\Pi(C))$ can be written as,
\begin{equation}
\begin{aligned}
\Pi &= \frac{1}{2}[\ln{\phi_{S}^{0}} + \ln{\phi_{C}^{0}}-\ln{(1-\phi_{P}-\phi_{C})}-\ln{\phi_{0}}+2\chi_{SC}{\phi_{C}^{0}}^{2} \\ 
 & +2\phi_{0}-2+2(1-\phi_{P}-\phi_{C})+2\frac{\phi_{P}}{N_{P}}+\chi_{PS}\phi_{P}(1-2\phi_{P}-2\phi_{C}) \\
 & +\chi_{PC}\phi_{P}(2\phi_{C}-1)-\chi_{SC}(2\phi_{P}\phi_{C}+2\phi_{C}^{2}-2\phi_{C}+1-\phi_{P})] .
\end{aligned}
\end{equation}
Then, second order virial coefficient can be obtained by expanding $\Pi$ about $\phi_{P}$ at $\phi_{P} \to 0$ and define the cosolvent fraction $x_{C}\equiv \frac{\phi_{C}^{0}}{\phi_{S}^{0}+\phi_{C}^{0}}$,
\begin{equation} \label{eq:7result_V2}
\begin{aligned}
\mathcal V_2 & = \frac{((\Delta\chi+\chi_{SC})^{2}-2\chi_{SC})x_{C}^{2}-2\Delta\chi-4\chi_{SC}+3}{4\chi_{SC}x_{C}^{2}-4\chi_{SC}x_{C}+2} + \frac{(-(\Delta\chi+\chi_{SC})^{2}+4\Delta\chi+6\chi_{SC})x_{C}}{4\chi_{SC}x_{C}^{2}-4\chi_{SC}x_{C}+2} \\
&  + \frac{2\chi_{SC}x_{C}-2\chi_{SC}+1}{(\lambda x_{C}+1)^{2}(4\chi_{SC}x_{C}^{2}-4\chi_{SC}x_{C}+2)} - \frac{(3\chi_{SC}+\Delta\chi)(x_{C}-1)+2}{(\lambda x_{C}+1)(2\chi_{SC}x_{C}^{2}-2\chi_{SC}x_{C}+1)} \\
& + \frac{1}{2} - \chi_{PS}
\end{aligned}
\end{equation}
where $\Delta\chi \equiv \chi_{PS}-\chi_{PC}$. The boundary of whether minimum, maximum or both can be observed can be found by taking the limit of the derivative of $\mathcal{V}_{2}$ with respect to $x_{C}$,
\begin{equation}
\lim_{x_{C} \to 0} \frac{\partial \mathcal{V}_{2}}{\partial x_{C}}=\chi_{SC}+\Delta\chi+\lambda-\chi_{SC}\lambda-\Delta\chi\lambda-\frac{(\chi_{SC}+\Delta\chi)^{2}}{2}
\end{equation} 
\begin{equation}
\begin{aligned}
\lim_{x_{C} \to 1} \frac{\partial \mathcal{V}_{2}}{\partial x_{C}} & = 2\Delta\chi-2\chi_{SC}-2\chi_{SC}\Delta\chi+\frac{(\chi_{SC}+\Delta\chi)^{2}}{2}+\frac{\chi_{SC}-\Delta\chi}{\lambda+1} \\
& +\frac{2\lambda}{(\lambda+1)^2}-\frac{\lambda}{(\lambda+1)^3} .
\end{aligned}
\end{equation}
In the $\chi_{SC} = 0$ case, four solutions can be derived by solving the obtained quadratic equation of $\Delta\chi$, 
\begin{equation}
\begin{aligned}
& \Delta\chi_{ctc}(B_{1})=(1-\lambda)+\sqrt{\lambda^{2}+1} \\
& \Delta\chi_{ctc}(B_{2})=-(1+\frac{\lambda}{1+\lambda})+\sqrt{4-\frac{8\lambda^{2}+9\lambda+3}{(\lambda+1)^{3}}} \\
& \Delta\chi_{ctc}(B_{3})=(1-\lambda)-\sqrt{\lambda^{2}+1} \\
& \Delta\chi_{ctc}(B_{4})=-(1+\frac{\lambda}{1+\lambda})-\sqrt{4-\frac{8\lambda^{2}+9\lambda+3}{(\lambda+1)^{3}}} .
\end{aligned}
\end{equation}

\subsection{Monte Carlo Simulation}
Single polymer chain with chain length equal to $50$ is immersed in solvents and cosolvents mixture. All of segments in the simulation are of equal size, with diameters being $\sigma$, which is the unit length in the simulation. The box size is $L_{x}=L_{y}=L_{z}=16 \sigma$. The total number of solvents and cosolvents is fixed at $n_{S}+n_{C} = 16334$ throughout all simulations, resulting in the overall number density of $4 / \sigma^{3}$. The respective value of $n_{S}$ and $n_{C}$ are adjusted to reflect the different solvent mixture composition. The simulation is performed in the NVT ensemble. In this setup, all polymer segments carry associating sites that can form reversible bonds with either solvents or cosolvents. However, associations between solvents and cosolvents are not permitted. Details on how associations are implemented in the Monte Carlo simulation can be found in the cited reference\cite{zhang2025investigation}. Setting $k_{B}T = 1$, the bonded potential for polymer chain can be written as,
\begin{equation} \label{eq:DGB_P}
u^{b}(\left | {\bf r}_{P,i} - {\bf r}_{P,j} \right |)=\frac{3 }{2a^{2}}\left | {\bf r}_{P,i} - {\bf r}_{P,j} \right |^{2}
\end{equation}
. And the non-boned interaction for each pair is,
\begin{equation}
u_{\alpha \alpha'}^{nb}({\bf r},{\bf r}') \equiv \epsilon_{\alpha\alpha'} \left( 15/2\pi \right) \left(1 - r/\sigma \right)^2
\end{equation}
, for $r<\sigma$, where $\sigma$ is the unit length, or $u_{\alpha \alpha'}^{nb}({\bf r},{\bf r}') = 0$ otherwise. $\epsilon_{\alpha\alpha'}$ controls the interaction strength and has the unit of $k_{B}T$, which is defined as, 
$\epsilon_{\alpha\alpha\prime}\equiv\left\{\def\arraystretch{1.2}\begin{tabular}{@{}l@{\quad}l@{}}
  $\epsilon_{\kappa}$ & if $\alpha=\alpha\prime$ \\
  $\epsilon_{\kappa}+\epsilon_{\chi_{\alpha \alpha'}}$ & if $\alpha \neq \alpha\prime$
\end{tabular}\right.$ . $\epsilon_{\kappa}$ is the excluded volume and $\epsilon_{\chi_{\alpha \alpha'}}$ describes the immiscibility. Conventional hopping trial moves are performed to equilibrate and sample the polymer configurations.\par

\par
\section{Results}
\subsection{Standard Flory-Huggins Theory - $\chi_{SC}$ Dominated Mechanism} \label{results:SFH_results}
Although standard, or non-association, Flory-Huggins theory is well established and applied to study polymer/solvent/cosolvent ternary mixture system \cite{dudowicz2015communication}, it is still important to discuss the parameter space curved out based on the second order virial coefficient ($\mathcal{V}_{2}$) behaviors. The way to calculate these boundaries can be found in the cited paper \cite{zhang2024general}. Cosolvency means the convex curve of $\mathcal{V}_{2}$ as a function of the cosolvent fraction, corresponding to globule-coil-globule transition, and, on the contrary, cononsolvency suggests the concave curve of $\mathcal{V}_{2}$, corresponding to coil-globule-coil transition. The "Co+Conon" region means that both a minimum and a maximum point can be observed in the $\mathcal{V}_{2}$ with respect ot $x_{C}$ plot. In figure~\ref{fig:7_figure_1.png}, $\chi_{SC}$ describes the interaction strength between solvents and cosolvents, and $\Delta\chi$ is defined as $\chi_{PS}-\chi_{PC}$, indicating the cosolvent excess interaction strength to polymer. The reason we use $\Delta\chi$ is that the boundary of different types of behaviors merely depends on relative strength of polymer-solvent and polymer-cosolvent interaction, not their absolute value. Cosolvency can only be observed when $\chi_{SC}$ is larger than $0$, corresponding to the repulsive interaction between solvents and cosolvents. The boundary for the occurrence of cosolvency is $\chi_{SC} - \Delta\chi > 0$ and $\chi_{SC}+\Delta\chi > 0$. It signifies that $\chi_{SC}$ should dominate the polymer interaction to solvent and cosolvent to observe cosolvency phenomenon, and that is the primary understanding for cosolvency phenomenon currently \cite{dudowicz2015communication,mukherji2017depleted}. When $\chi_{SC}$ is equal to $0$ or negative, no cosolvency can be observed. As it is already discussed in the introduction, more and more evidence suggest that the picture of cosolvency induced by $\chi_{SC}$ is incomplete.\par

\Figure{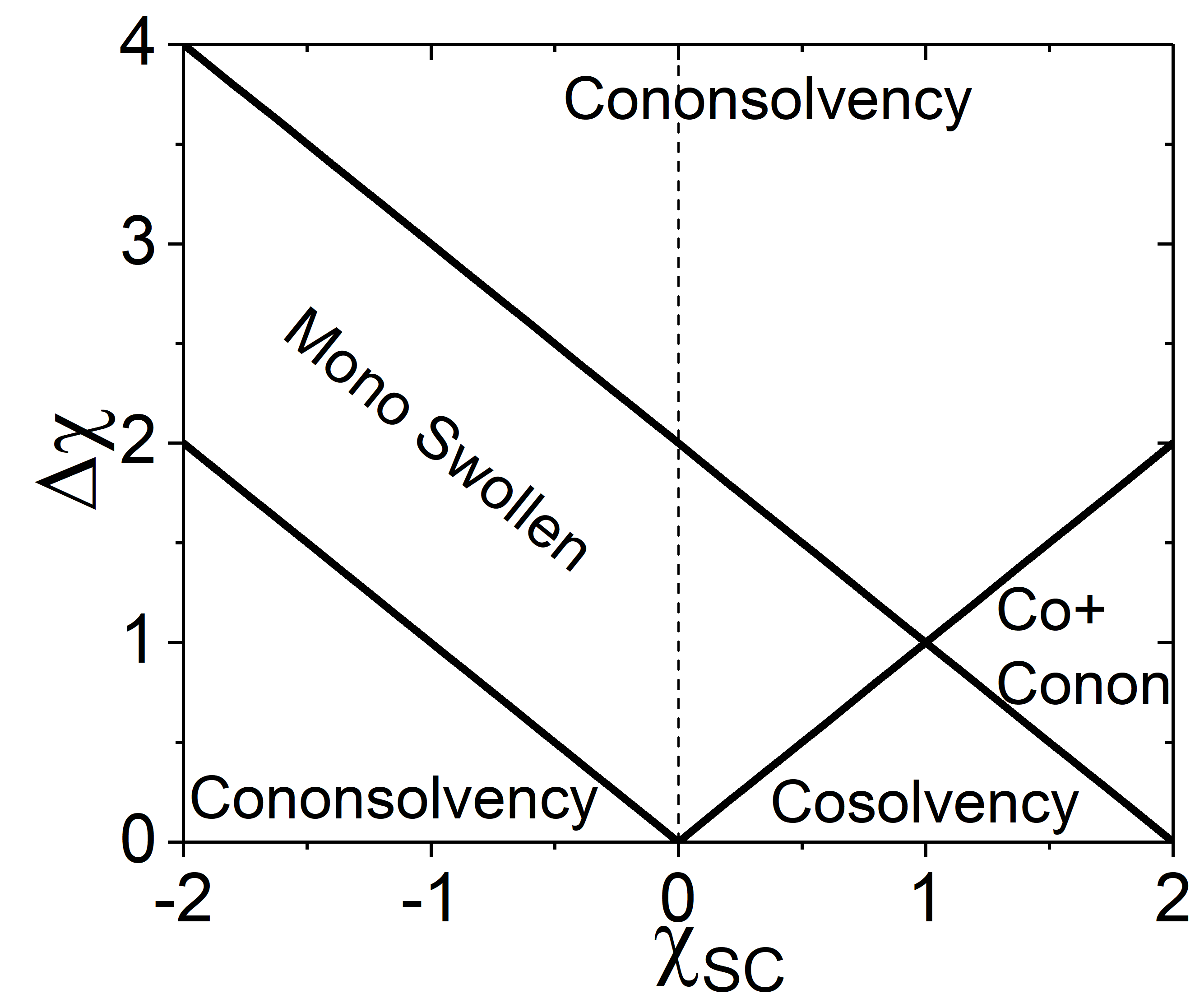}{0.7\linewidth}{Parameter space of $\Delta\chi$ and $\chi_{SC}$ calculated in the framework of non-associative Flory-Huggins theory.}

\subsection{The Effect of Association Predicted by Flory-Huggins Theory}\label{results:7result_AFH_results}
Based on the reasoning from experiments, hydrogen bonds play a certain role in cosolvency phenomenon \cite{hoogenboom2008tuning, lambermont2010temperature}. So, association is introduced in Flory-Huggins theory to obtain a more comprehensive picture. In our model, only cosolvent has the ability to associative with the polymer. The reason we set this up is based on previous work, in which the feature recovers with the non-associative Flory-Huggins theory when both solvents and cosolvents have the same equilibrium constant \cite{dudowicz2015phase}. So, according to their results, it is safe to say that only the relative difference of P-C and P-S association strength is important, similar to $\Delta\chi$ effect in non-association Flory-Huggins theory. \par
First, we examine a second order virial coefficient example plot in figure~\ref{fig:7_figure_2.png} calculated based on associative Flory-Huggins theory with $\chi_{SC}$ being equal to $0$. $x_{C}$ is the cosolvent fraction. 
In this system, $\chi_{PS}$ is equal to $1$, corresponding to a bad solvent condition. $\chi_{PC}$ is equal to $2.18$, and $\lambda$ is $5$. Hence, $\Delta\chi$ is equal to $-1.18$. 
As it is well known, $\mathcal{V}_{2}$ being equal to $0$ suggests the theta solvent for the polymer \cite{rubinstein2003polymer}. With $\mathcal{V}_{2}$ being larger zero, polymer exhibits extended state, and with it being less than zero, polymer is in collapsed state \cite{rubinstein2003polymer}. At two end points, corresponding to pure solvents or pure cosolvents system, polymer exhibits collapsed state. With the addition of the cosolvent, the polymer transits to swollen state, and $\mathcal{V}_{2}$ goes back to the negative region if $x_{C}$ is further increased. This is a typical cosolvency behavior of $\mathcal{V}_{2}$, and it shows the similar trend as former experimental reports, signifying the validity of investigating cosolvency based on associative F-H theory \cite{horta1981preferential,katime1984polymer}. \par

\Figure{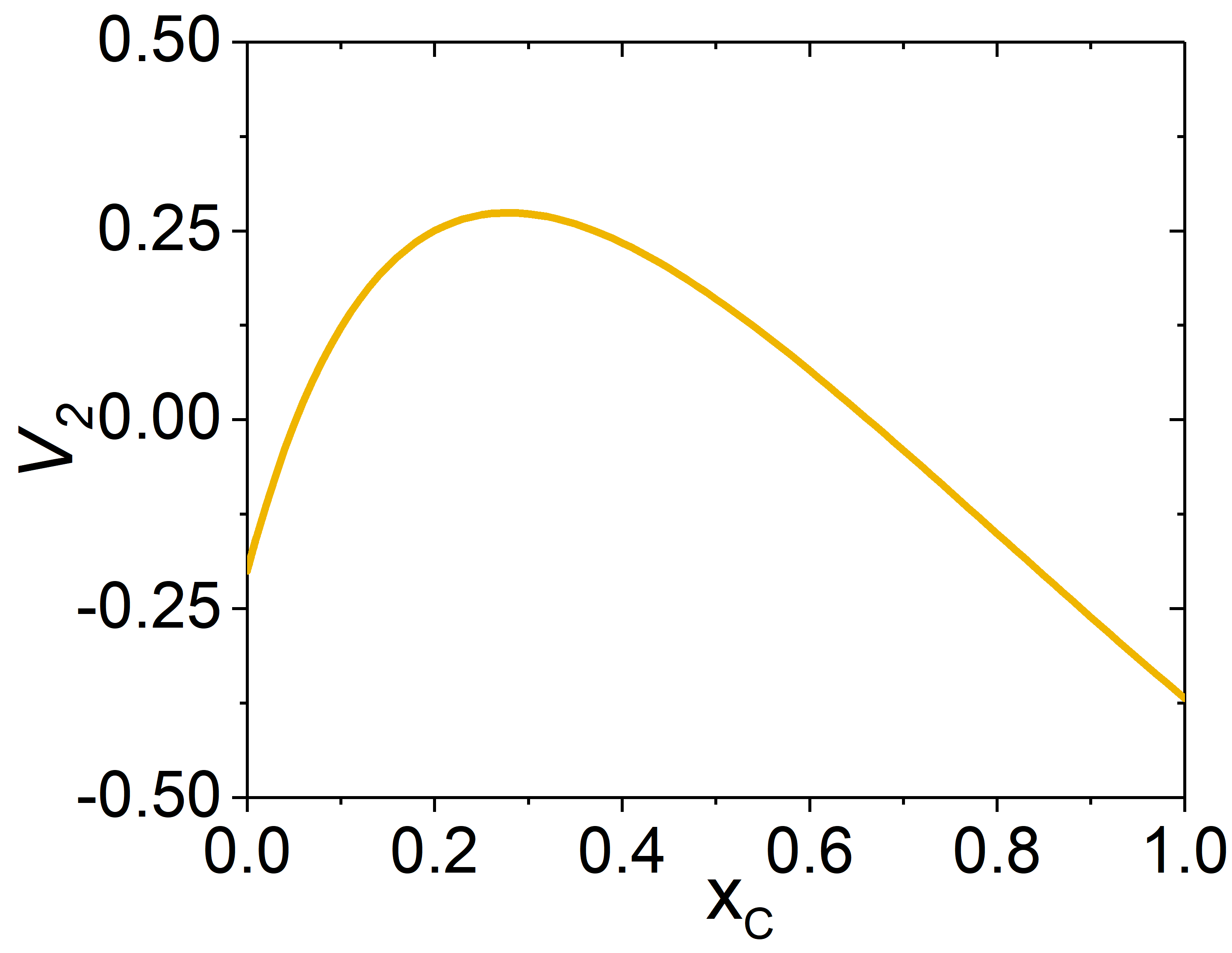}{0.7\linewidth}{Second order virial coefficient of P/S/C system with $\chi_{PS}=1, \chi_{PC}=2.18, \chi_{SC}=0, \lambda=5$ plotted against cosolvent fraction.}

Next, in figure~\ref{fig:7_figure_3.png}, the parameter space of $\Delta\chi$ with respect to $\lambda$ at $\chi_{SC}=-1$, $\chi_{SC}=0$ and $\chi_{SC}=1$ is shown, which is based on how $\mathcal{V}_{2}$ behaviors will change with the increase of the cosolvent fraction. For example, "monotonic swollen" region means that the $\mathcal{V}_{2}$ keeps increasing with the addition of the cosolvent within that parameter setting, and, correspondingly, "monotonic collapse" region means the $\mathcal{V}_{2}$ decreases with the addition of the cosolvent. $\lambda$ is the association strength, implying P-C association equilibrium constant, and $\Delta\chi$ is defined as $\chi_{PS}-\chi_{PC}$. The major difference with the parameter space predicted by standard Flory-Huggins theory is the removal of requiring $\chi_{SC}>0$ being the prerequisite for the occurrence of cosolvency, indicating that solvent-cosolvent interaction becomes less important. And cosolvency window (including "cononsolvency+cosolvency" region) is expanded evidently with the increase of the $\chi_{SC}$ value, from $\chi_{SC}=-1$ to $1$, indicating that $\chi_{SC}$ may play the supporting role in the occurrence of cosolvency phenomenon. Thereby, $\chi_{SC}$ is set to be zero in our calculations to screen out $\chi_{SC}$ effect, as it is not the deciding factor.\par

\Figure{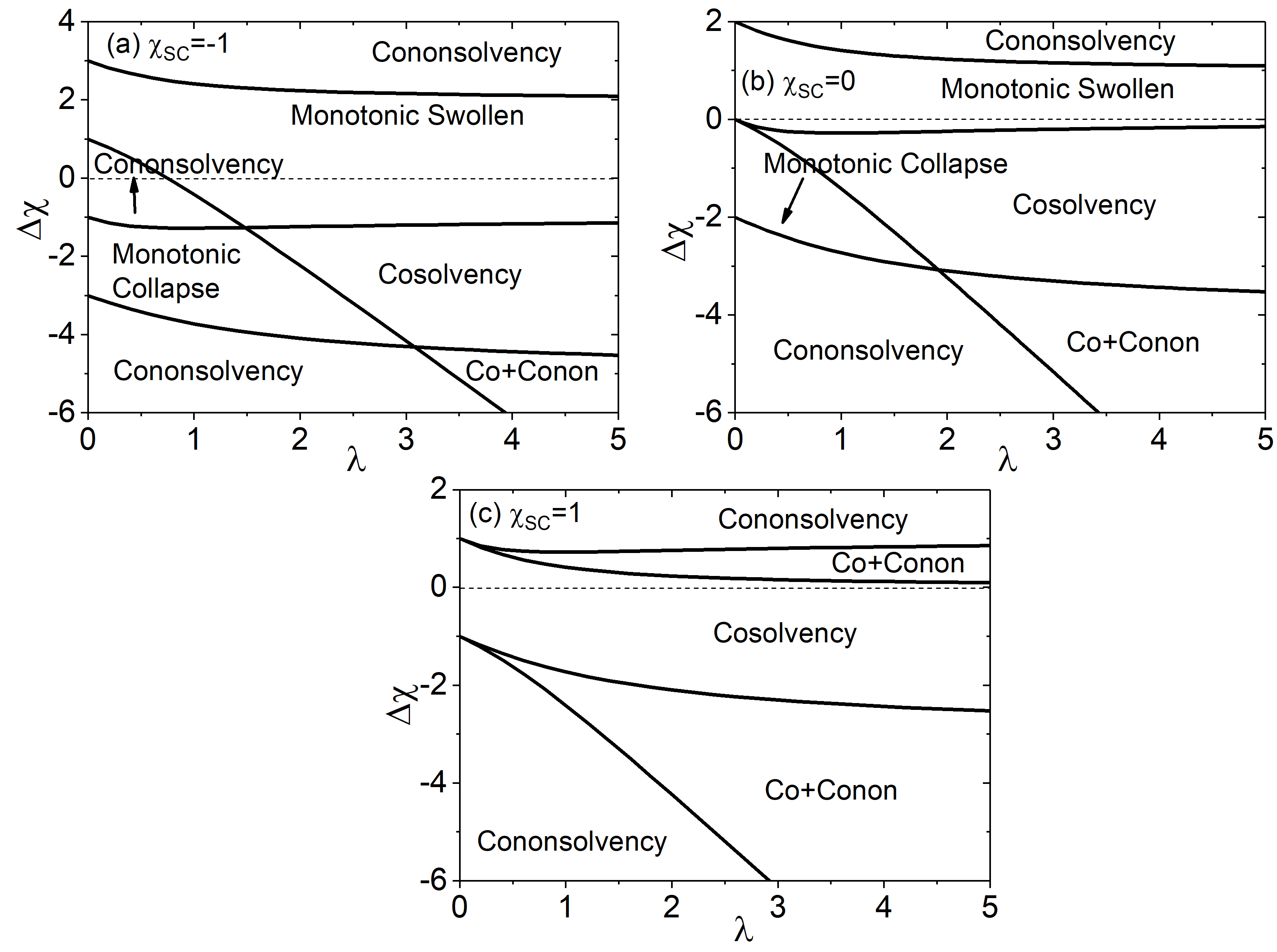}{0.98\linewidth}{Parameter space of $\Delta\chi - \lambda$ calculated by associative Flory-Huggins theory at (a) $\chi_{SC}=-1$, (b) $\chi_{SC}=0$ and (c) $\chi_{SC}=1$, of which the region is divided based on $\mathcal{V}_{2}$ behaviors.}

Based on parameter space calculated by associative Flory-Huggins theory when $\chi_{SC}=0$, the occurrence of cosolvency requires the negative $\Delta\chi$, suggesting that polymer-cosolvent should have stronger repulsive interactions than polymer-solvent, and $\lambda \neq 0$, indicating that there should be the extent of association between polymer and cosolvent. Generally speaking, polymer-solvent and polymer-cosolvent should have different association tendency and repulsive interactions.\par

\subsection{Monte Carlo Simulation Examination} \label{results:MC_results}
To justify associative F-H predictions and obtain more details about cosolvency phenomenon, coarse-grained Monte Carlo simulation is applied. It is not practical to scan all over the parameter space, so, we pick three representative points to run the simulation with $\chi_{SC}=0$ like showing in figure~\ref{fig:7_figure_4.png} (a). In the first system, polymer-cosolvent have stronger repulsive interaction than polymer-solvent ($\epsilon_{\chi_{PC}}=0.9, \epsilon_{\chi_{PS}}=0.6$), corresponding to VDW type interaction asymmetry, and with the same association tendency for P-C and P-S (${h_{A}}_{S}={h_{A}}_{C}=3$). The second system has association tendency asymmetry (${h_{A}}_{S}=3, {h_{A}}_{C}=0$), but with same VDW type interactions to the polymer ($\epsilon_{\chi_{PC}}=\epsilon_{\chi_{PS}}=0.9$). The third system is chosen within the cosolvency region, with both VDW type interaction difference ($\epsilon_{\chi_{PC}}=1, \epsilon_{\chi_{PS}}=0.35$) and association tendency asymmetry (${h_{A}}_{S}=6, {h_{A}}_{C}=1$). One thing needed to mention is that the smaller the $h_{A}$ value is, the stronger the association tendency. End-to-end distance ($R_{e}^{2}$) is plotted against cosolvents fraction in figure~\ref{fig:7_figure_4.png} (b). In the first and second system, $R_{e}^{2}$ monotonically changes with the addition of the cosolvents. And a maximum point of $R_{e}^{2}$ can be observed in third system, signifying the occurrence of cosolvency phenomenon. Correspondingly, conversion of solvents and cosolvents is plotted against cosolvent fractions in figure~\ref{fig:7_figure_5.png}. It is defined as the number of attached S or C segments on the polymer chain divided by the total chain length. For both the 2nd and 3rd systems, the association of cosolvents overwhelms the association of solvents. But in the first system, it only shows the linear change.\par

\Figure{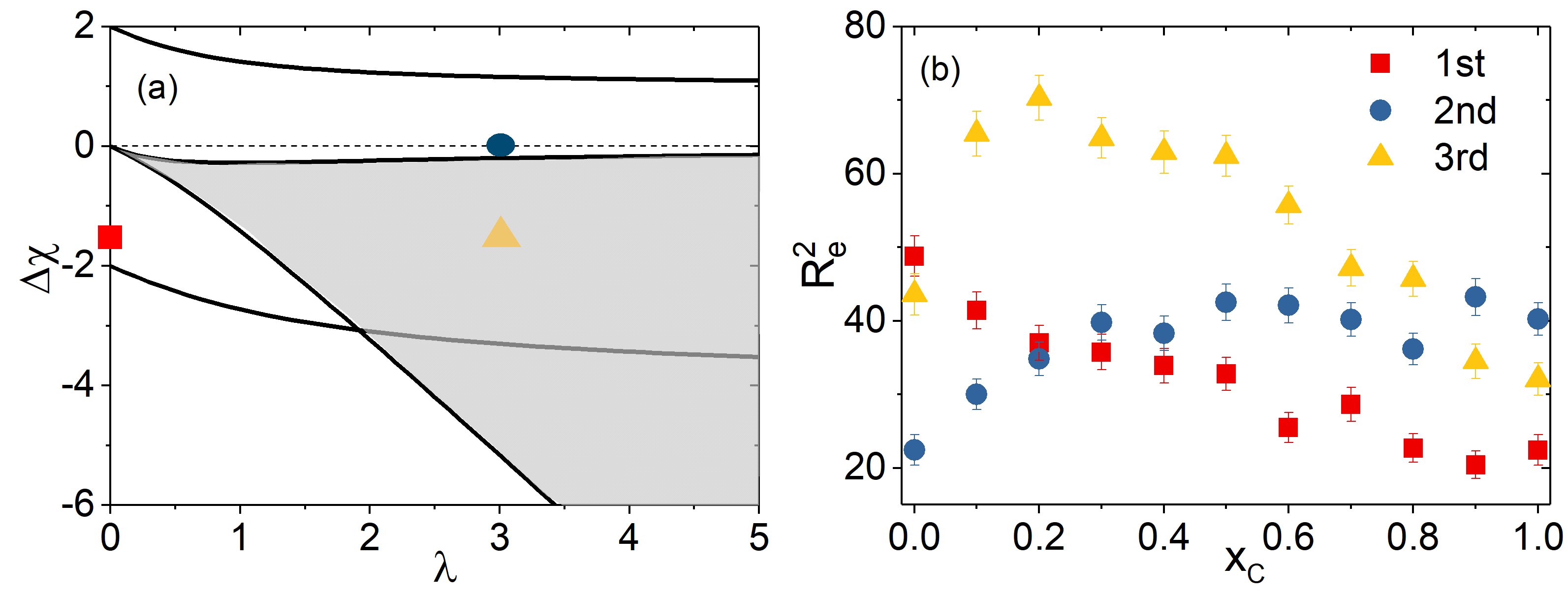}{0.98\linewidth}{(a) Three set of parameters are chosen to run MC simulation based on associative FH prediction, 1st: $\epsilon_{\chi_{PC}}=0.9, \epsilon_{\chi_{PS}}=0.6, {h_{A}}_{S}={h_{A}}_{C}=3$; 2nd: $\epsilon_{\chi_{PC}}=\epsilon_{\chi_{PS}}=0.9, {h_{A}}_{S}=3, {h_{A}}_{C}=0$; 3rd: $\epsilon_{\chi_{PC}}=1, \epsilon_{\chi_{PS}}=0.35, {h_{A}}_{S}=6, {h_{A}}_{C}=1$. (b) End-to-end distance is plotted against cosolvent fraction in three MC simulations.}

The reason for the first system conformation behavior is straightforward. The added cosolvent is the worse solvent than the primary solvent for the polymer. And they have the same associating probability, thereby, linear change of solvent and cosolvent associating rate is observed in figure~\ref{fig:7_figure_5.png}. Consequently, polymer chain tends to become more collapsed with the replacement of the primary solvent by the worse secondary solvent. That is similar to most of system exhibiting monotonic change with the addition of the non-solvent, like poly[oligo(ethylene glycol) methyl ether methacrylate] (POEGMA) immersed in isopropanol/hexane system \cite{roth2011ucst}. In the second system, polymer chain becomes more extended with the addition of the cosolvent. Cosolvent has larger associating tendency with polymer than solvents, but they have the same repulsive interaction to the polymer. So, the associating rate at $x_{C}=1$ is higher than it at $x_{C}=0$. The only change brought about by the addition of the cosolvent is the number of attached segments on the polymer chain. Accordingly, the chain will become more swollen with the increase of $x_{C}$ due to purely steric effects. In above two systems, no cosolvency can be observed. And the polymer chain conformation characterized by MC simulation perfectly fits the parameter space predicted by associative Flory-Huggins theory.\par

\Figure{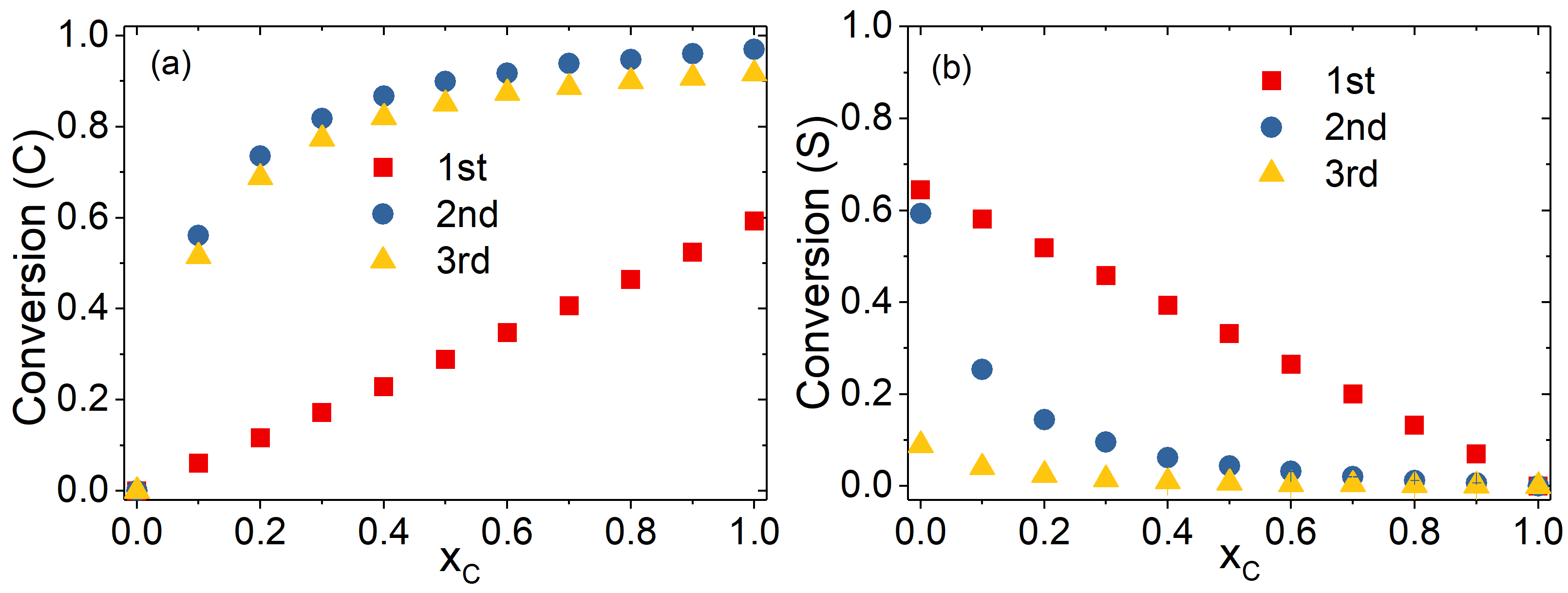}{0.98\linewidth}{Conversion of (a) cosolvents  and (b) solvents plotted against cosolvent fraction in three MC simulation system.}

In the third system, an evident maximum point at $x_{C}=0.2$ can be observed. $R_{e}^{2}$ behavior shows globule-coil-globule transition, exhibiting typical cosolvency behaviors. In figure~\ref{fig:7_figure_5.png}, both second and third system show the higher association rate of cosolvents than solvents. But there are two interesting points here. First, only third system exhibits cosolvency. The second system shows monotonic conformation change. Second, the maximum point is at $x_{C}=0.2$, but that is not the point where polymer chain is most saturated. So, the radial distribution of the cosolvents and solvents around the polymer segment is plotted at $x_{C}=0.2$ system in figure~\ref{fig:7_figure_6.png} to obtain more details. The enrichment of cosolvents around the polymer segment can be observed for both second and third system, and it is even more enriched in second system, as it is shown in figure~\ref{fig:7_figure_6.png} (a). Correspondingly, figure~\ref{fig:7_figure_6.png} (b) shows a slight degree of depletion of solvents in second system, but not for third system. Thereby, it indicates that only cosolvent decoration is not enough for the occurrence of cosolvency phenomenon. Combined with results from three types of systems, it can be conjectured that cosolvency is actually the synergy effect of non-bonded VDW type interactions and associative interactions. In pure solvents, polymer exhibits collapsed state as solvent environment is bad. Next, the addition of the cosolvent with stronger association tendency than solvents will cause the enrichment of cosolvents inside the polymer coil like showing in figure~\ref{fig:7_figure_6.png} (a). The attached cosolvents on the polymer chain try to expand the polymer chain due to P-C strong repulsive interactions, corresponding to red force in figure~\ref{fig:7_figure_6.png} (c). And the surrounding segments exert the force to collapse the polymer chain, corresponding to blue force drawn in figure~\ref{fig:7_figure_6.png} (c). When solvent/cosolent fraction reaches a certain mixing ratio, the swollen force will overwhelm the collapse force due to VDW type interaction difference, and thereby, the chain tends to be expanded. In the recovering process, when most of solvents are replaced by cosolvents, the collapse force applied by surrounding segments will also be increased because of P-C has stronger repulsive interaction than P-S. In conclusion, cosolvency phenomenon results from the composited nature of VDW type interactions and associative interactions. It requires cosolvent has both stronger association tendency and VDW type repulsive interaction than solvents.\par

\Figure{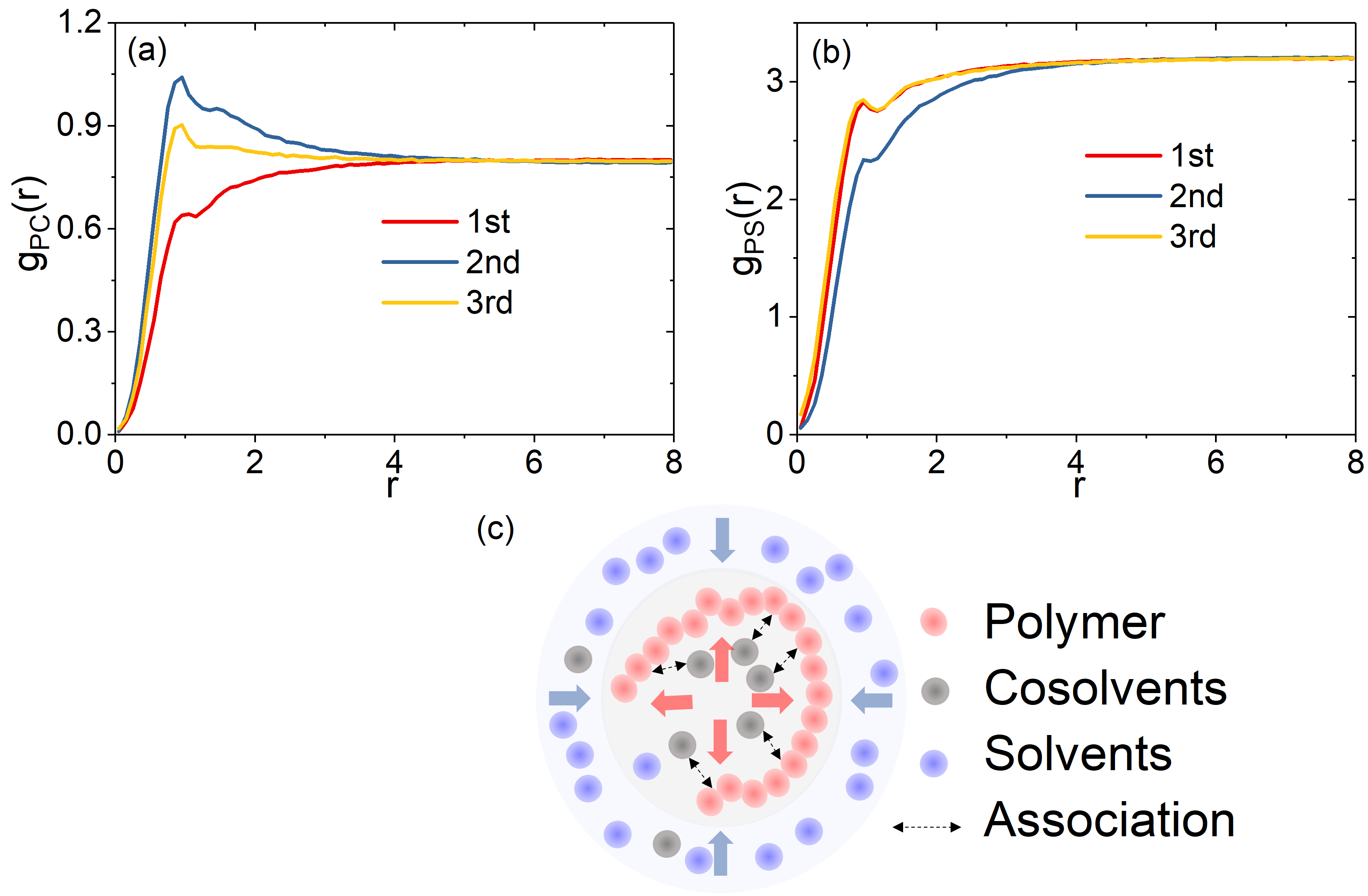}{0.98\linewidth}{Radial distribution of (a) cosolvents and (b) solvents around polymer segments in the first, second and third system. (c) The illustration of the competition of the repulsive force between polymer - attached cosolvents and polymer - surrounding solvents.}
 
\subsection{Experimental Results Comparison} \label{7results:Exp_results}
The extent of polymer conformation swollen in MC simulation, which is the third system, is compared with previous experimental results \cite{lee2014enhanced}. In figure~\ref{fig:7_figure_7.png} (a), $\sqrt{R_{e}^{2}(x_{C})/R_{e}^{2}(x_{C}=1)}$ is plotted for the third system. And hydrodynamic diameter (HD) normalized by HD at $x_{C}=1$ is plotted for PMMA (molecular weight=$15000$) immersed in 2-propanol/water system at $318.15K$ \cite{lee2014enhanced}. The degree of swollen in MC simulation is similar to experimental results, and the maximum position are both at $x_{C}=0.2$ point. Afterwards, phase diagram calculated by associative Flory-Huggins theory at the same set of parameters as MC simulation is also compared with the same experimental system. Roughly, given similar association rate, the $\lambda$ parameter is set as $3$ for cosolvents, and $\lambda=\infty$ for solvent system, meaning no association. The MC parameter for the third system simulation can be converted to the F-H parameter based on the following equation:
\begin{equation}
\chi_{ij}-\chi_{\theta}=\rho_{0}(\epsilon_{\chi_{ij}}-\epsilon_{\chi_{\theta}})
\end{equation} 
where $\chi_{\theta}$ is the $\theta$ point at given $\lambda$, for $\lambda=\infty$, $\chi_{\theta}$ is just $0.5$, for $\lambda=3$, $\chi_{\theta}=1.53$. $\rho_{0}$ is the overall density in the simulation, which is $4$, and $\epsilon_{\chi_{\theta}}$ is the point where chain conformation exhibits ideal behavior, equivalently,  $\left\langle R_{e}^{2} \right\rangle =R_{e}^{2}(ideal)$. When $h_{A}=1$, $\epsilon_{\chi_{\theta}}=0.7$, when $h_{A}=6$, $\epsilon_{\chi_{\theta}}=0.3$. Based on the above relation, it can be found out that $\lambda_{C}=3, \chi_{PC}=2.73$, and $\lambda_{S}=\infty, \chi_{PS}=0.7$, so, that will be the parameter used to calculate the phase diagram. \par
Figure~\ref{fig:7_figure_7.png} (b) shows the ternary phase diagram of experimental results and F-H predictions \cite{lee2014enhanced}. F-H theory and experimental results both show two separated demixing region, and the dexming area boundary predicted by F-H theory is close to the experimental measurement. Hence, the degree of swollen of the polymer chain measured by MC simulation and the phase diagram calculated by associative F-H theory given a same set of parameters are both comparable with experimental data. \par

\Figure{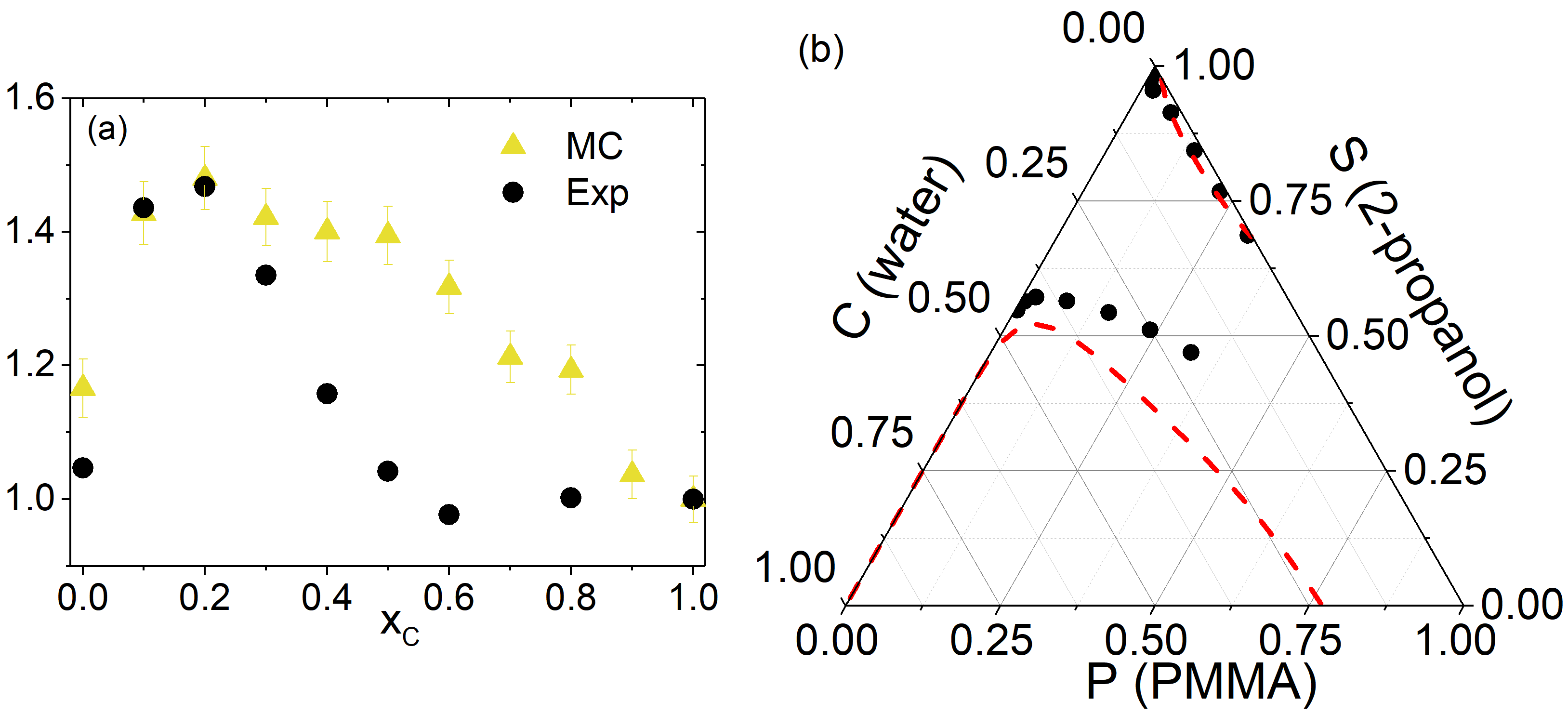}{0.98\linewidth}{(a) The comparison of degree of swollen for MC simulation ($\sqrt{R_{e}^{2}(x_{C})/R_{e}^{2}(x_{C}=1)}$) and experiments (hydrodynamic diameter ($x_{C}$)/hydrodynamic diameter ($x_{C}=1$)). (b) The comparison of ternary phase diagram calculated from associative Flory-Huggins theory (dashed lines) and measured from experiments (sphere symbols).}

\section{Discussion}
According to former experimental reports, the enrichment of cosolvents around the polymer can be observed in cosolvency systems \cite{katime1984polymer,horta1981preferential,prolongo1984cosolvency, cowie1972polymer,viras1988preferential}. So, it indeed indicates some extents of preferential adsorption of cosolvents around the polymer. Additionally, the increased solubility of 2-phenyl-2-oxazoline (PhOx) in ethanol is reported with the addition of the water, and it has been proposed that the increased solubility arises from the formation of hydration shell around the carbonyl group \cite{hoogenboom2008tuning}. This is similar to PMMA/alcohol/water systems, as water tends to form the hydrogen bond to the ester moieties of PMMA \cite{zhang2015polymers}. Another evidence is that copolymers of oligo(ethylene glycol) methyl ether acrylate and oligo(ethylene glycol) phenyl ether acrylate (p(OEGMeA-co-OEGPhA)) show increased solubility at intermediate water/ethanol composition, and polymer prefer hydrogen bonding with water than ethanol \cite{roth2013advancing}. In conclusion, the occurrence of cosolvency accompanies with the preferential hydrogen bonding of polymer with cosolvents. As it has been proposed before, the hydration shell acts as the compatibilizing layer to dissolve polymer \cite{hoogenboom2008tuning, lambermont2010temperature, zhang2015polymers}. But that is not the only prerequisite for cosolvency. In the second system of MC simulation, cosolvents are preferentially adsorbed on the polymer chain like showing in figure~\ref{fig:7_figure_5.png} (a) and figure~\ref{fig:7_figure_6.png} (a), but no cosolvency can be observed, consistent with associative F-H prediction, which is in monotonic swollen region. We attribute this to the lack of VDW type interactions difference. \par
In PMMA brush immersed in ethanol/water and isopropanol/water system, PMMA brush height exhibits nonmonotonic change, and the maximum height of PMMA brush is higher in water/2-propanol mixture than ethanol/water mixture \cite{yu2015cosolvency}. We can attribute this to the difference of 2-propanol and ethanol polarity, which in fact corresponds to $\chi_{PS}$ value difference. The order according to polarity for them is, water$>$ethanol$>$2-propanol, so ,generally speaking, the repulsive interaction strength of PMMA to them can be written as, PMMA-water$>$PMMA-ethanol$>$PMMA-2-propanol. Therefore, the VDW type interaction difference in 2-propanol/water mixture is larger than ethanol/water mixture. To be more comprehensible, the blue force illustrated in figure~\ref{fig:7_figure_6.png} (c) is weaker in 2-propanol/water mixture than ethanol/water mixture, but with the same strength of red force. Accordingly, PMMA brush is more extended in 2-propanol/water mixture. \par
In conclusion, the effect of cosolvents compatibilization should be clarified, and the way how cosolvency comes into the picture actually needs more factors according to our results. Because the cosolvent has the stronger repulsive interactions with polymer than that of solvents with polymers, polymer will try to minimize the contact area with cosolvents, and to dissolve in solvents. Additionally, the composited nature of enhanced solvation is not only fit for macromolecules, but also for small molecules. A recent report reveals the cosolvency mechanism for tolbutamide in alcohol/toluene mixture \cite{li2022revealing}. In their study, the preferential hydrogen bond of solute with alcohol can be observed. So, the high association tendency of alcohol with solute results in the enrichment of alcohol around the solute, and intramolecular interaction is undermined not only due to the competition of hydrogen bonding sites but also the attached alcohol repulsive interaction to surrounding solutes. Therefore, the solute would rather be dissolved than aggregated because of the composited competition of solute/solvents/cosolvents association tendency and VDW type interactions. \par
Based on the above mechanism, it is straightforward to explain the different solvent/cosolvent ratios to give the maximum solubility among all cosolvency systems \cite{ikkai2003swelling,boyko2003preparation,lee2014enhanced}. As the cosolvency is a composited effect of VDW type interaction difference and associative interaction asymmetry, different combinations of solvent/cosolvent pairs and polymer types will give different $\Delta\chi$ (VDW type interaction difference) and $\Delta\lambda$ (association tendency asymmetry) value. Therefore, the maximum of the solubility plot can be at any points. Similar argument has been made before, suggesting that the mixing ratio in which gives the maximum miscibility depends on species polarity \cite{zhang2015polymers}. At last, a general property of solvent/cosolvent mixture, like excess volume, cannot be found among all types of cosolvency solvent/cosolvent pairs \cite{villamanan1985excess,lama1965excess,cowie1972polymer,shalmashi2014densities,tahery2006density}. So, solvent-cosolvent interactions may not be so important, or at least, cannot decide the occurrence of cosolvency phenomenon. The effect of solvents-cosolvents interaction is well illustrated by associative F-H theory parameter space in figure~\ref{fig:7_figure_2.png}. The larger the $\chi_{SC}$ is, the more evident of cosolvency phenomenon. And the negative $\chi_{SC}$ value, corresponding to strong attractive interaction between solvents and cosolvents, will impair the strength of cosolvency. \par

\newpage
\bibliographystyle{unsrt}
\bibliography{CS_bulk_citation}

\end{document}